\documentclass[sigconf]{acmart}

\usepackage{soul}
\usepackage{xcolor}
\usepackage{float}
\usepackage{subcaption}
\usepackage{makecell}

\AtBeginDocument{%
  }

\copyrightyear{2025}
\acmYear{2025}
\setcopyright{acmlicensed}
\setcctype{by}
\acmConference[MM '25]{Proceedings of the 33rd ACM International Conference on Multimedia}{October 27--31, 2025}{Dublin, Ireland}
\acmBooktitle{Proceedings of the 33rd ACM International Conference on Multimedia (MM '25), October 27--31, 2025, Dublin, Ireland}\acmDOI{10.1145/3746027.3754593}
\acmISBN{979-8-4007-2035-2/2025/10}

\begin{document}


\title[SimViews]{SimViews: An Interactive Multi-Agent System Simulating Visitor-to-Visitor Conversational Patterns to Present Diverse Perspectives of Artifacts in Virtual Museums}


\author{Mingyang Su}
\authornote{Both authors contributed equally to this research.}
\affiliation{%
  \institution{The Hong Kong University of Science and Technology (Guangzhou)}
  \city{Guangzhou}
  \country{China}
}
\email{sumy22@mails.tsinghua.edu.cn}

\author{Chao Liu}
\authornotemark[1]
\affiliation{%
  \institution{The Hong Kong University of Science and Technology (Guangzhou)}
  \city{Guangzhou}
  \country{China}
}
\email{cliu009@connect.hkust-gz.edu.cn}

\author{Jingling Zhang}
\affiliation{%
  \institution{The Hong Kong University of Science and Technology (Guangzhou)}
  \city{Guangzhou}
  \country{China}
}
\email{jzhang898@connect.hkust-gz.edu.cn}

\author{WU Shuang}
\affiliation{%
  \institution{The Hong Kong University of Science and Technology (Guangzhou)}
  \city{Guangzhou}
  \country{China}
}
\email{shuang@u.nus.edu}

\author{Mingming Fan}
\authornote{Corresponding author}
\affiliation{%
  \institution{The Hong Kong University of Science and Technology (Guangzhou)}
  \city{Guangzhou}
  \country{China}
}
\affiliation{%
  \institution{The Hong Kong University of Science and Technology}
  \city{Hong Kong}
  \country{China}
}
\email{mingmingfan@ust.hk}

\renewcommand{\shortauthors}{Mingyang Su, Chao Liu, Jingling Zhang, WU Shuang and Mingming Fan}

\begin{abstract}
Offering diverse perspectives on a museum artifact can deepen visitors’ understanding and help avoid the cognitive limitations of a single narrative, ultimately enhancing their overall experience. Physical museums promote diversity through visitor interactions. However, it remains a challenge to present multiple voices appropriately while attracting and sustaining a visitor’s attention in the virtual museum. Inspired by recent studies that show the effectiveness of LLM-powered multi-agents in presenting different opinions about an event, we propose SimViews, an interactive multi-agent system that simulates visitor-to-visitor conversational patterns to promote the presentation of diverse perspectives. The system employs LLM-powered multi-agents that simulate virtual visitors with different professional identities, providing diverse interpretations of artifacts. Additionally, we constructed 4 conversational patterns between users and agents to simulate visitor interactions. We conducted a within-subject study with 20 participants, comparing SimViews to a traditional single-agent condition. Our results show that SimViews effectively facilitates the presentation of diverse perspectives through conversations, enhancing participants' understanding of viewpoints and engagement within the virtual museum.
\end{abstract}




\begin{teaserfigure}
  \centering
  \includegraphics[width=1\textwidth]{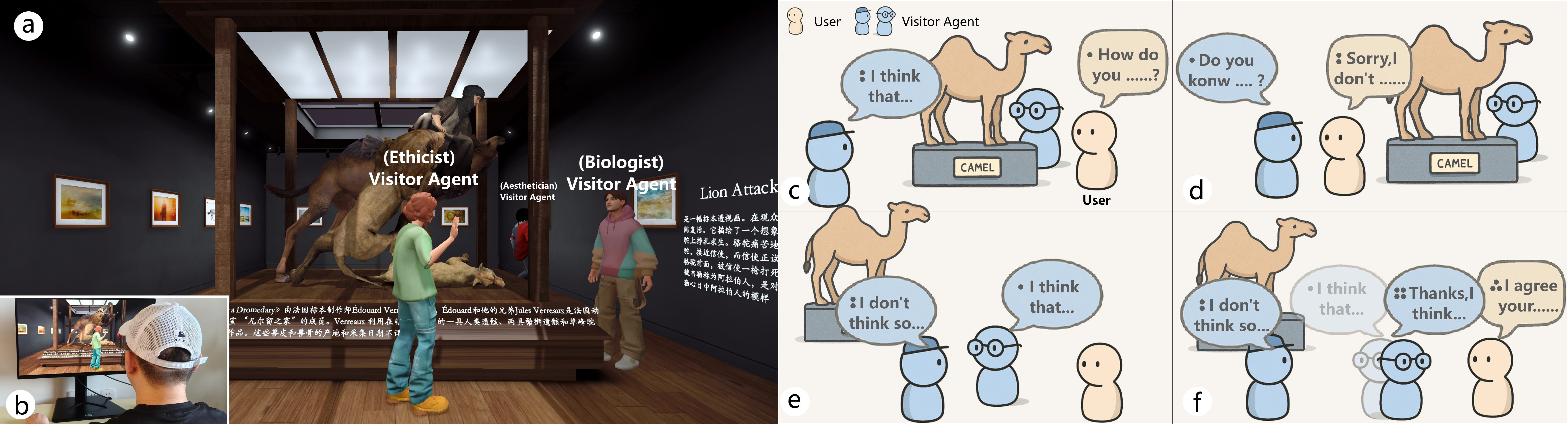}
  \caption{(a) Multi-agent conversation in the virtual museum to present diverse perspectives; (b) The participant engages in a conversation with the visitor agent; Four types of conversation patterns: (c) \textbf{The user's active speaking}: the user actively initiates a conversation with the agent; (d) \textbf{The user's passive speaking}:  the agent initiates a conversation with the user, and the user responds; (e) \textbf{The user's passive listening}: the user overhears the agents' conversation; (f) \textbf{The user's active listening}: the user listens to the agents' conversation, and they join in if they want.}
  \label{fig:title}
\end{teaserfigure}

\maketitle
\section{Introduction}
A long-standing metaphor for the museum is a ‘forum’ for different voices to debate and converse \cite{cameron1972museum,karp1991exhibiting}. Presenting diverse perspectives in museums can help visitors move beyond personal biases, fostering more inclusive and meaningful experiences \cite{damiano2022exploring,sleeter2008making,chin2013key}. Specifically, it can deepen understanding \cite{chung2009presenting,chu2019embodied,reynolds2011reinventing}, encourage cultural reflection \cite{abidin2024exploring}, and enhance engagement \cite{falk2016museum,natusch2015role}. In physical museums, such diversity is typically promoted through participatory practices, such as inviting underrepresented groups to contribute to exhibition narratives during curation \cite{lavine1991exhibiting}, or collecting visitor feedback through comment books and cards \cite{dallas2004presence,nashashibi2003visitor}. 

With the rapid advancement of digital technology, virtual museums are currently contributing to cultural diversity. However, these diversities focus more on the content of exhibits, purposes of usage, and access \cite{gran2019digital}, ignoring the presentation of diverse perspectives within the museum. Researchers suggest the use of AI practice to achieve this goal and provide a more inclusive service \cite{huang2022social}.

Recent studies have shown that Large Language Models (LLMs) can effectively simulate agents within a given domain \cite{li2023camel,qian2023communicative}. These agents are widely used in social media \cite{zhang2024see}, online decision-making \cite{park2023choicemates}, strategic consulting \cite{neves2024squad}, and other areas to provide diverse voices through dialogue, which may offer the potential to display diverse perspectives in virtual museums. However,
most of these are primarily text-based, the museum should actively encourage and facilitate different forms of conversation \cite{saerkjaer2024visitor}, rather than being confined to a static and monologic wall text. 

Inspired by these works, we explored \emph{how to design an LLM-powered multi-agent conversation method to effectively present diverse perspectives through conversation with users (\textbf{RQ1})}. To this end, we developed SimViews, an interactive system that simulates visitor-to-visitor conversational patterns to support the presentation of diverse perspectives in virtual museums. The system uses LLMs to create visitor agents with distinct professional identities based on literature \cite{zhang2024see}, and structures 4 visitor-to-visitor conversation patterns between users and agents by drawing on Goffman’s participation framework \cite{goffman1981forms}: \textbf{active speaking} (user initiates), \textbf{passive speaking} (agent initiates), \textbf{passive listening} (user overhears agent conversations), and \textbf{active listening} (user joins the ongoing discussion of two agents), as shown in Fig.~\ref{fig:title}.

We also explored \emph{whether and how multi-agent conversations supported by SimViews may help users understand these diverse perspectives and engage more deeply with the museum (\textbf{RQ2})}. To address this, we implemented a prototype of a virtual museum featuring two exhibits and six agents. We then conducted a user study with 20 participants. The results showed that \textsc{simviews} effectively facilitates the presentation of diverse perspectives through conversations, enhancing participants' understanding of viewpoints and engagement within the virtual museum. Among the four patterns, participants rated \textbf{active listening} as most helpful for understanding and engagement, while \textbf{active speaking} was preferred for gaining control.  Based on these findings, we present design insights for presenting diverse perspectives in a virtual museum through multi-agent conversation design. Overall, our work made the following contributions:

\begin{itemize}
    \item  We propose a multi-agent system that simulates 4 visitor-to-visitor conversational patterns to present diverse perspectives in virtual museums.
    \item We implemented a functional prototype and conducted a user study to evaluate the design, offering design insights into presenting multiple perspectives in virtual museums using multi-agent systems.
\end{itemize}

\section{Related Work}
\subsection{Diverse Museum}
To embrace and foster cultural diversity, ICOM redefined museums as "democratizing, inclusive, and polyphonic spaces" \cite{ICOM2021}, encouraging the inclusion of diverse voices from outside the institution in exhibit interpretation \cite{reynolds2011reinventing}.In physical museums, such diversity is increasingly realized through curatorial practices that involve underrepresented groups, particularly minority communities, which sometimes includes seeking interpretations from the communities who created, used, or are associated with them \cite{lavine1991exhibiting, srinivasan2010diverse}.Visitor participation also plays a key role in presenting multiple perspectives during museum visits \cite{reynolds2011reinventing}. Many museums now employ interactive installations \cite{nashashibi2003visitor} and multimedia technologies \cite{porsche2018public} to support this goal. For example, Tate Britain’s Turner Prize exhibition includes a space for visitors to share their responses, though these are often limited to text-based displays. At Randers Art Museum, visitors were invited to comment on artworks, and their reflections were later exhibited, fostering perspective diversity through interaction and dialogue \cite{saerkjaer2024visitor}.

Virtual museums also contribute to cultural diversity, primarily by broadening access to diverse audiences and offering a wide range of content \cite{wang2024virtuwander, gran2019digital}. However, this form of diversity often overlooks the presentation of multiple perspectives on the artifacts themselves, which remains a significant gap in the virtual museum experience.

\subsection{LLM-powered Multi-agent}
LLMs have demonstrated remarkable capabilities in simulating domain-specific agents through coherent prompting and in-context learning \cite{qian2023communicative, li2023camel}. By maintaining consistent personas and delivering context-aware responses, LLMs can effectively embody expert roles over extended interactions \cite{liu2023pre, xu2023expertprompting}. Building on this capability, recent studies have introduced LLM-based agents into virtual museum environments, where they serve as personalized tour guides. These agents can offer professional, adaptive explanations and contextual guidance tailored to users’ interests \cite{wang2024virtuwander, trichopoulos2023large2, trichopoulos2023large}. 

To further enrich interaction, researchers have explored multi-agent frameworks that leverage LLMs' reasoning through inter-agent collaboration. Multi-agent setups allow agents to discuss, debate, and converge on more accurate or nuanced answers \cite{guo2024large, yu2023co, zheng2023chatgpt, du2023improving}. For instance, multiple expert agents can jointly analyze medical cases to reach a diagnostic consensus \cite{tang2023medagents}. Beyond accuracy, multi-agent systems also introduce identity diversity, agents can be designed to represent different expertise, personalities, or viewpoints, thereby reducing the uniformity of responses typically generated by a single LLM \cite{liang2023encouraging, briesch2023large}. One example is Zhang et al.'s system that simulates diverse agents with distinct professional backgrounds during social media browsing to expose users to a broader range of perspectives \cite{zhang2024see}.

\subsection{Agent Conversation}
Benefiting from LLMs' powerful understanding and generation capabilities, multi-agents can engage in more natural interactions with humans. However, in current studies, user interactions with multi-agents are still primarily based on dialogue boxes \cite{engelmann2023maids, zhang2024see, tang2023medagents}. For example, the Finch financial advisory system enables users to make decisions through text-based conversations with four collaborative agents \cite{de2018specifying, de2018ravel}. While it showcases the potential of multi-agent collaboration in complex tasks, it offers limited immersion and interactivity in virtual environments.

To improve user experience, recent studies have explored virtual avatars and voice interaction to support more natural conversations \cite{traum2002embodied, liu2024classmeta, liu2025toward}. For instance, HUMAINE introduced voice interaction in simulated transaction negotiations \cite{divekar2020humaine}, but its agents only engage in dialogue with users, lacking inter-agent discussions. Agent United provides a multi-agent dialogue platform, where although agents are equipped with virtual avatars, users can only participate through text-based response options \cite{beinema2021agents}, limiting both flexibility and realism. These examples highlight that achieving truly natural and flexible multi-agent conversation remains a significant research challenge.

\section{System Design}

To address RQ1, we proposed key design considerations and developed a multi-agent system that incorporates visitor agent identities and structured conversation patterns to support inclusive and engaging visitor interactions.
\subsection{Agent Design}
 For agent design, we addressed two primary considerations: \textbf{(D1.1) Design virtual agents with distinct professional identities}. Museum visitors often have multiple social identities, leading to diverse interpretations of exhibits \cite{sawyer2013studying}. For example, historians viewing a Bengal tiger specimen from India in an English museum might interpret it as "a cipher for hegemonic Victorian attitudes toward India and its people" \cite{crane2014picturing}, whereas animal conservationists could emphasize ethical objections to hunting practices \cite{green2006tiger}. To reflect this interpretive diversity, our agents were created with specific professional identities instead of generic visitor personas.
\textbf{(D1.2) Equip agents with multimodal representations, including visual avatars and anthropomorphic voices}. Museums need interactive presentations beyond static, text-based displays to effectively engage visitors \cite{saerkjaer2024visitor}. Prior studies highlight that multimodal forms of information significantly enhance participant engagement and comprehension in virtual environments \cite{raptis2021mumia, lin2019multimodal}. Additionally, agents with visual avatars and natural voices are more persuasive and engaging \cite{baylor2011design}. Therefore, our agent representations include detailed avatar models and voice interactions instead of mere textual dialogue boxes.

\subsection{Conversation Design}

For conversation design, our goal is to simulate the visitor-to-visitor conversational patterns. Inspired by recent studies that have applied Goffman's participation framework \cite{goffman1981forms, dynel2011revisiting} to multi-agent dialogue settings \cite{samson2020two, wang2013multi}, we adopt this framework to guide our design and identified two key dimensions for structuring visitor-to-visitor conversational patterns: \textbf{(D2.1) Define user roles as speaker or listener}. In multi-agent environments, users may act as speakers, addressed listeners, or bystanders. Differentiating these roles allows for the design of varied and realistic conversational interactions.\textbf{ (D2.2) Specify whether the user or the agent initiates the conversation}. Clarifying conversational initiative helps determine interaction flow and user engagement levels, guiding how agents are scripted to respond or lead.

\subsection{Design Features}

\subsubsection{Professionally Grounded Agent Identities.} To address the agent identity requirement (\textbf{D1.1}), we assigned each visitor agent a distinct professional identity based on the scholarly perspectives of exhibits, such as ethicists, art historians, and biologists. Prior studies have shown that simulating conversations across professional roles enables users to access more diverse perspectives \cite{zhang2024see}. However, unlike open-ended online platforms, museums demand more professional and accurate representations \cite{Neill2008museums, Gaynor2005museum}. To reduce hallucinations and ensure credibility, each agent’s identity and dialogue were grounded in scholarly literature, using content authored by domain experts. This design enables the system to present reliably diverse and academically sound interpretations of the exhibits.

\subsubsection{Embodied and Multimodal Agent Representation.} 
\begin{figure}[h]
    \centering
    \includegraphics[width=1\linewidth]{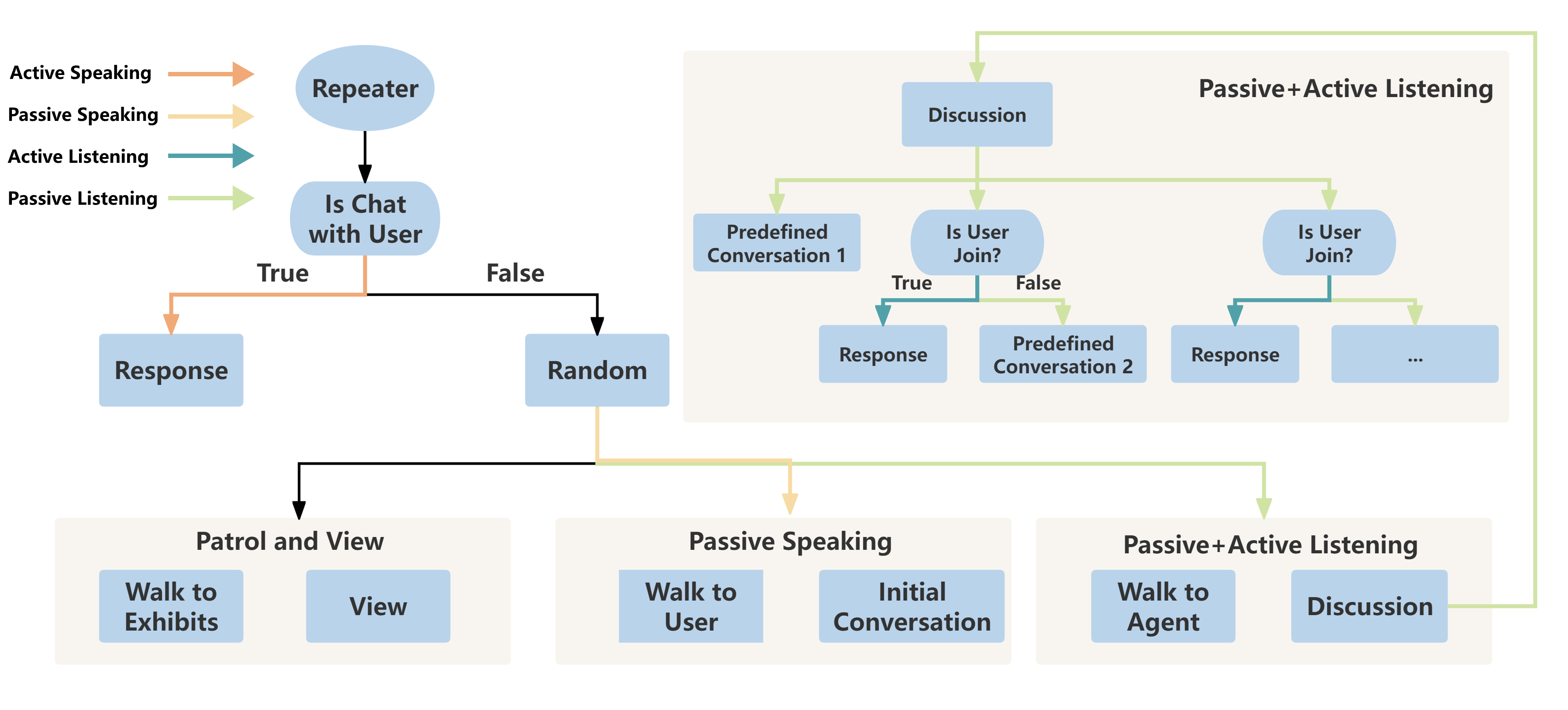}
    \caption{Agent behavior tree for switching between patrol, viewing, and four conversational patterns.}
    \label{fig:tree}
\end{figure}

To fulfill representation considerations (\textbf{D1.2}), we provided each agent with a full-body 3D avatar and a synthesized voice to facilitate natural voice-based interactions. This multimodal embodiment goes beyond traditional dialogue boxes and enhances immersion. Research has shown that voice and visual embodiment increase believability, trust, and engagement in virtual agents \cite{baylor2011design, raptis2021mumia}. Identity labels are hidden by default and shown only when users interact with an agent, helping participants distinguish agent roles while minimizing cognitive load and visual clutter.  To simulate lifelike museum dynamics, agents autonomously navigate the gallery, observe exhibits, and interact based on behavior tree logic. As shown in Fig. \ref{fig:tree}, when a user enters, agents exhibit socially present behaviors such as walking, gesturing, and conversing.

\subsubsection{Multi-Pattern Conversational Framework.} 

Based on the identified conversation dimensions (\textbf{D2.1} and \textbf{D2.2}), we proposed a 2×2 matrix to categorize different patterns of user-agent conversations. This structure yields four conversation types: \textbf{Active speaking}: The user initiates the conversation, and the agent responds. This is the most common interaction pattern in traditional user-agent systems and virtual environments, such as when a visitor asks a guide agent a question \cite{wang2024virtuwander}.\textbf{ Passive speaking}: The agent initiates the conversation, and the user responds. This reflects recent attention to agent proactivity and system-driven engagement strategies \cite{liu2020speaker}. \textbf{Active listening}: The user listens to an ongoing conversation between agents and may choose to join. This corresponds to an eavesdropping-like behavior where the user transitions from passive observer to active participant. \textbf{Passive listening}: the user overhears an agent conversation without actively engaging. This mode represents ambient awareness or unintended reception of dialogue, aligning with Goffman's notion of overhearers.

\section{SimViews Prototype Implementation}
Building on our design features, we developed \textbf{\textsc{simviews}} as a working prototype that brings multi-agent visitor conversations into a virtual environment.

\subsection{Virtual Museum Setup}
\begin{figure}[h]
    \centering
    \includegraphics[width=1\linewidth]{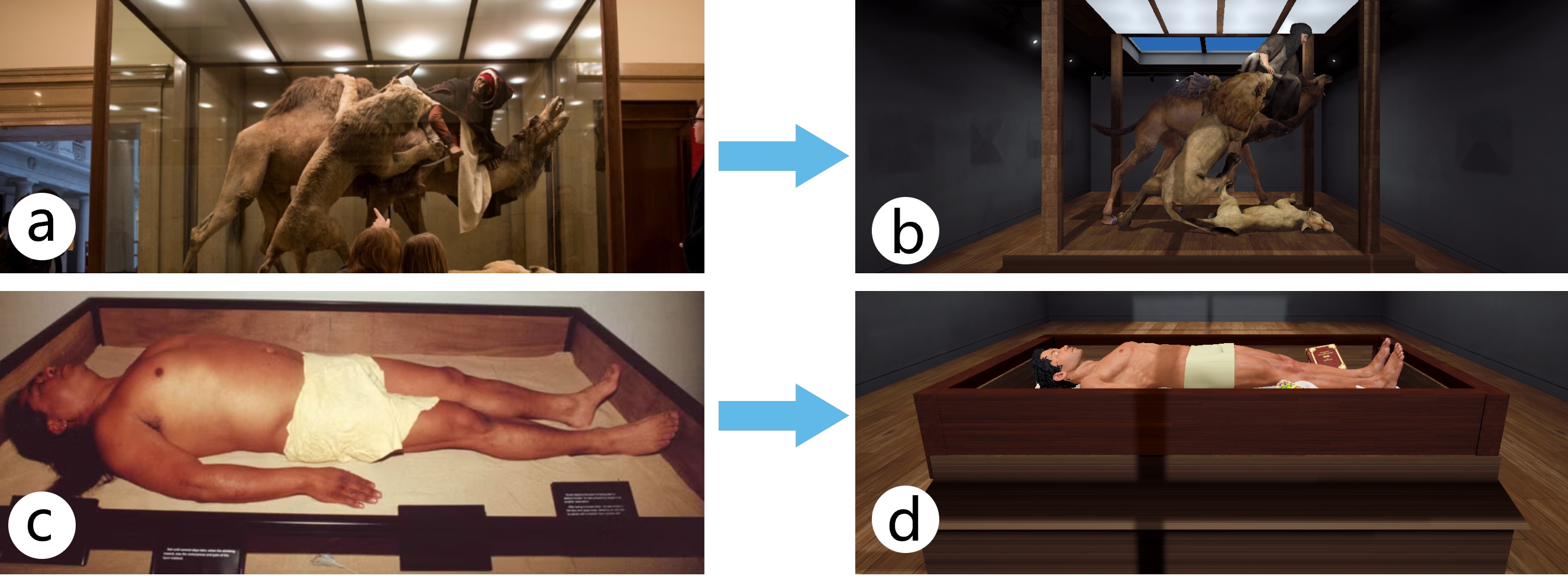}
\caption{Original and reconstructed versions of Lion Attacking a Dromedary (a–b) and The Artifact Piece (c–d).}
    \label{fig:exhibit}
\end{figure}

We built a virtual gallery in Unity and selected two exhibits with contrasting interpretations from prior literature: the sculpture \emph{Lion Attacking a Dromedary} and the performance artwork \emph{The Artifact Piece}. The former depicts a dramatic animal attack \cite{niittynen2023lions}, while the latter is a self-exhibition by Native American artist James Luna challenging ethnographic display norms \cite{hawley2016james}. These exhibits were selected to maximize interpretive contrast, thematic richness, and scholarly coverage. Both have been widely discussed from multiple disciplinary perspectives, making them suitable anchors for multi-agent conversations. To ensure visual fidelity, we invited an animator to reconstruct both exhibits in 3D based on photographs. Lighting, textures, and materials were carefully refined, and exhibit descriptions were displayed within the environment (see Fig.~\ref{fig:exhibit}).

\subsection{Constructing Multiple Visitor Agents}

We extracted three distinct viewpoints for each exhibit from relevant literature and assigned them to separate agents to ensure a clear and coherent expression of diverse perspectives. For \textit{Lion Attacking a Dromedary}, the viewpoints include: (1) an emphasis on the sculpture’s Romantic narrative and visual drama; (2) a critique of the use of human remains, raising concerns about racial violence; and (3) a challenge to the depiction of nature as inherently violent, advocating for a more balanced ecological perspective \cite{niittynen2023lions, fellous2023natural, black2013examining}. These viewpoints were assigned to three different agents. Then, we assigned corresponding identity labels to these agents based on these viewpoints, in this case \textbf{Aesthetician} , \textbf{Ethicist}, and \textbf{Biologist}. For \textit{The Artifact Piece}, we adopted the same approach and assigned the extracted perspectives to \textbf{Art Historian}, \textbf{Indigenous Scholar}, and \textbf{Curator}, based on their literature \cite{blocker2001failures, blocker2009ambivalent, lauzon2013playing, hawley2016james}.

Each agent features a unique appearance, identity, and conversational capability. For agent representation, avatars were generated using Ready Player Me, with randomized attributes (e.g., gender, skin tone, and clothing). To reduce stereotypes, agent appearance was not linked to specific professions. Identity labels were hidden until interaction began, ensuring that participants focused on the agents' conversational content rather than visual cues. Character animations such as standing, walking, talking, and thinking were integrated using Unity’s Animator. To introduce variability, Unity’s Behavior Tree and Animator were used to randomize these actions. Each agent’s voice was synthesized using Azure’s text-to-speech service, matching the avatar gender.Lip-sync was implemented via BlendShapes and synchronized with speech output.

\subsection{Multi-agent conversational flow}
\begin{figure}[H]
    \centering
    \includegraphics[width=1\linewidth]{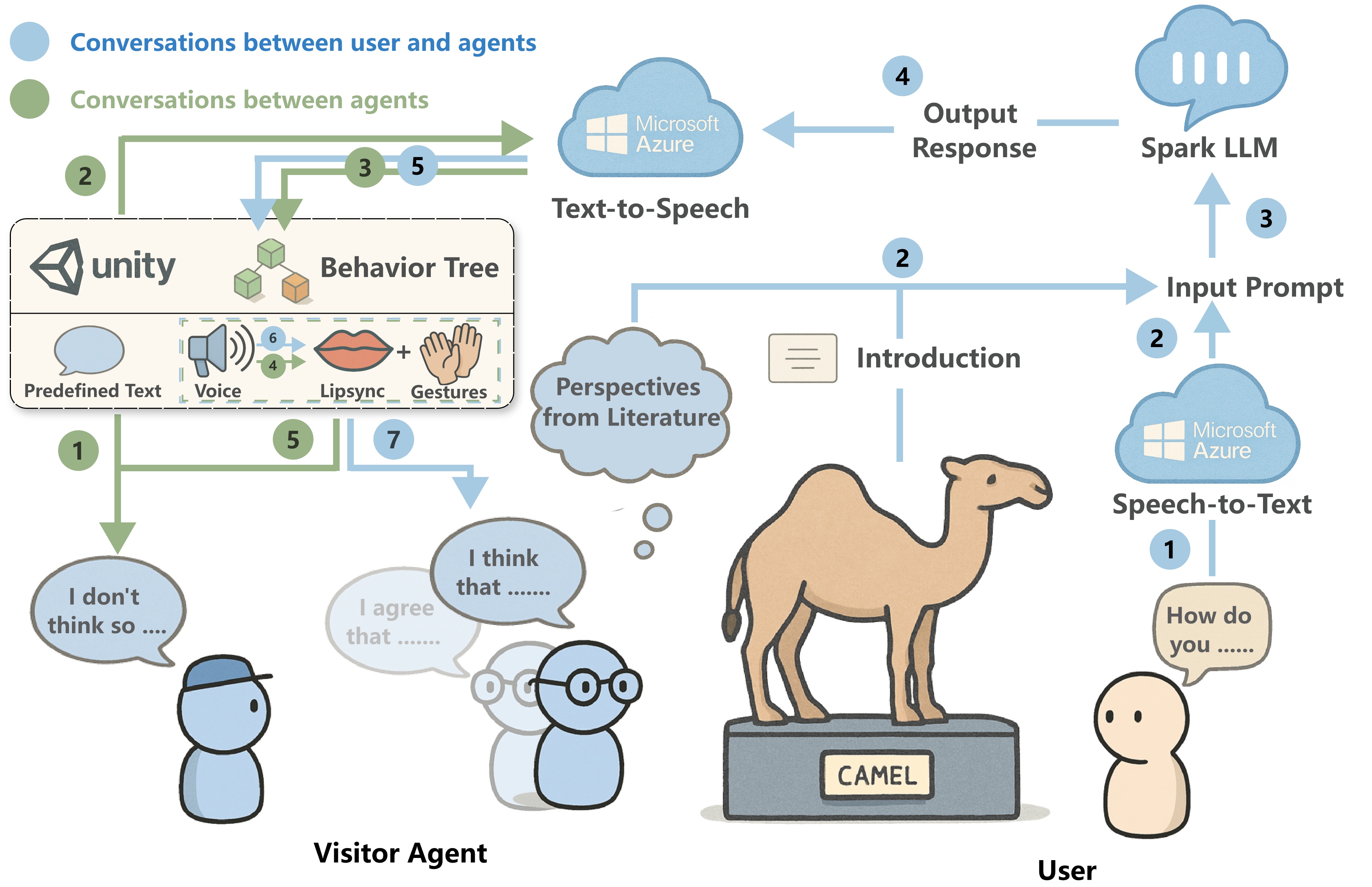}
    \caption{Interaction flow of a multi-agent system supporting voice-based user-agent and agent-agent conversations.}
    \label{fig:systemoverview}
\end{figure}

The system integrates agent-agent and user-agent conversation flows within a real-time virtual environment (see Fig. \ref{fig:systemoverview}).
To enable autonomous multi-agent interaction, each agent is programmed to initiate or respond to spoken interaction based on role and proximity. Agent-to-agent interactions are driven by predefined scripts drawn from literature to present contrasting perspectives. These conversations are coordinated through behavior trees, as shown in Fig. \ref{fig:tree}, ensuring narrative coherence and accuracy. Agents also autonomously perform actions such as walking, observing exhibits, initiating conversations, or engaging in discussion. When interacting with users, the system transcribes speech using Azure’s speech-to-text service, combines it with exhibit context and literature-based viewpoints, and sends it as a prompt to the Spark LLM. The generated response is then synthesized into speech using Azure TTS, with lip movement and hand gestures automatically synchronized. These mechanisms create a lifelike and socially engaging environment that simulates visitor-to-visitor conversations in physical museums.

\section{USER STUDY}

\subsection{Study Design}

\begin{figure}[h]
\centering
\includegraphics[width=1\linewidth]{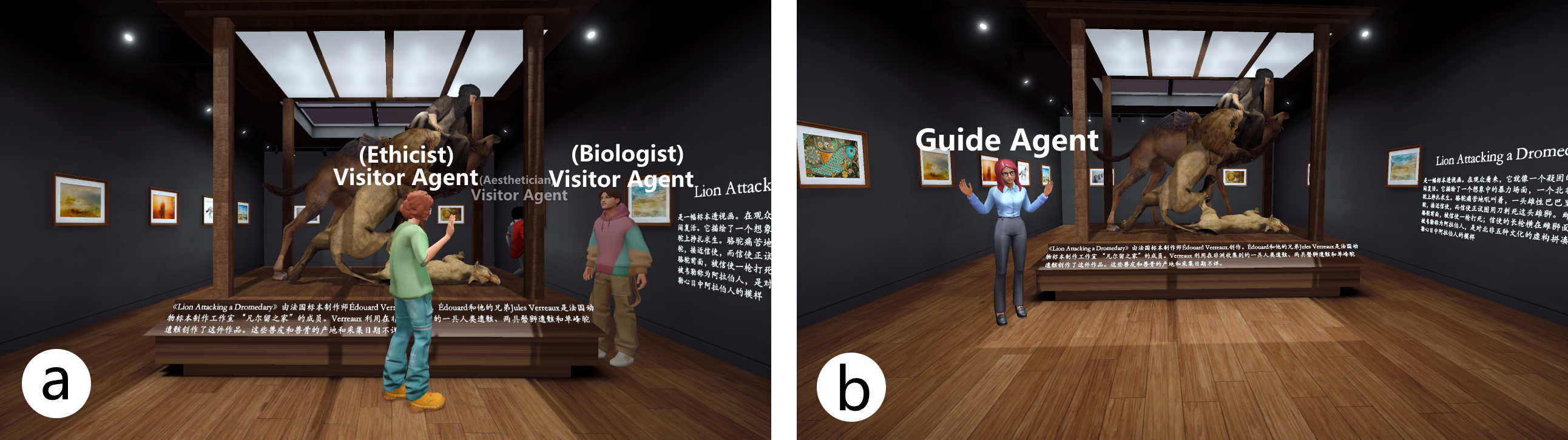}
\caption{(a) In \textbf{\textsc{simviews}}: participants interact with visitor agents; (b) In \textbf{\textsc{base}}: participants listen to the guide agent or ask questions.}
\label{fig:condition}
\end{figure}

To address RQ2, we conducted a within-subject comparative experiment. The experiment included two conditions, illustrated in Fig.~\ref{fig:condition}. In the \textbf{\textsc{simviews}} condition, participants interacted with multiple agents, each representing a distinct expert identity. In the \textbf{\textsc{base}} condition, participants interacted with a single guide agent, which presented all viewpoints in a unified narrative, sequentially explaining perspectives based on the same literature sources as those used in \textbf{\textsc{simviews}}. Participants could ask follow-up questions and receive direct responses. Each participant experienced both conditions with different exhibits, and the order was counterbalanced using a Latin Square design. This study was reviewed and approved by the ethics committee at the first author’s institution.

\subsection{Participants}
We recruited 20 participants (M=25.15, SD=2.43, aged 18-34) through word-of-mouth and snowball sampling. To minimize bias in understanding diverse perspectives, we recorded their professional backgrounds and museum visit frequency. Participants represented diverse disciplines: 4 from Computer Science, 3 from Environmental Studies, 3 from Design, 2 each from Economics, Materials Science, Biomedical Sciences, and Social Sciences, and 1 each from Chemistry and Intelligent Manufacturing. In terms of visit habits, 40\% visited semiannually, followed by 30\% quarterly, and 15\% each for monthly and annual visits.

\subsection{Procedure}

Before the study began, participants completed a practice session using the Leeds Tiger exhibit, presented in a separate gallery. This exhibit, not used in the main study, allowed participants to familiarize themselves with the interaction modes of both \textbf{\textsc{simviews}} and \textbf{\textsc{base}}, with researcher guidance provided throughout. All interactions were recorded. Each participant spent at least 10 minutes in each exhibit, and their behavior was recorded. After each condition, participants completed a knowledge test, shared their viewpoints, and filled out a questionnaire. Researchers followed up with brief questions and semi-structured interviews to gather deeper insights and design feedback.  

To evaluate our systems, we collected four types of data: custom questionnaires, conversational data, knowledge tests, and viewpoint statements. Questionnaires captured participants’ subjective experiences. Conversational data recorded user-agent interactions and were coded for dialogue frequency and type. Knowledge tests, consisting of 10 researcher-designed multiple-choice questions, and viewpoint statements were used to evaluate participants’ understanding of diverse perspectives. 

\section{Results}

\subsection{Custom Questionnaire}

\begin{figure*}[h]
    \centering
\includegraphics[width=1\linewidth]{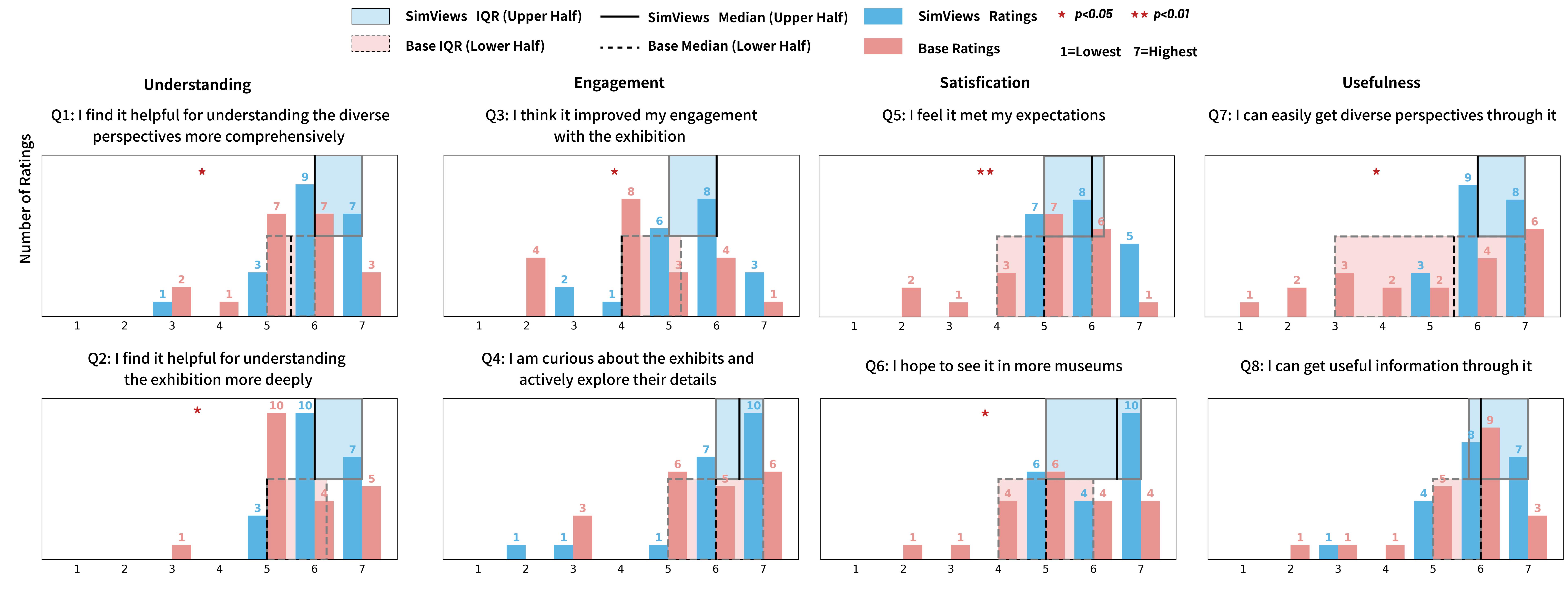}
 \caption{Results of our custom questionnaire on a Likert scale (1 = strongly disagree, 7 = strongly agree). Blue represents ratings for \textbf{\textsc{simviews}}, while red represents ratings for \textbf{\textsc{base}}. The shaded areas indicate the interquartile range (IQR), and the vertical lines mark the median values for each group. Statistical significance is indicated as *p < 0.05, ** p < 0.01.}
    \label{fig:res}
\end{figure*}

\begin{figure}[h]
    \centering
    \includegraphics[width=1\linewidth]{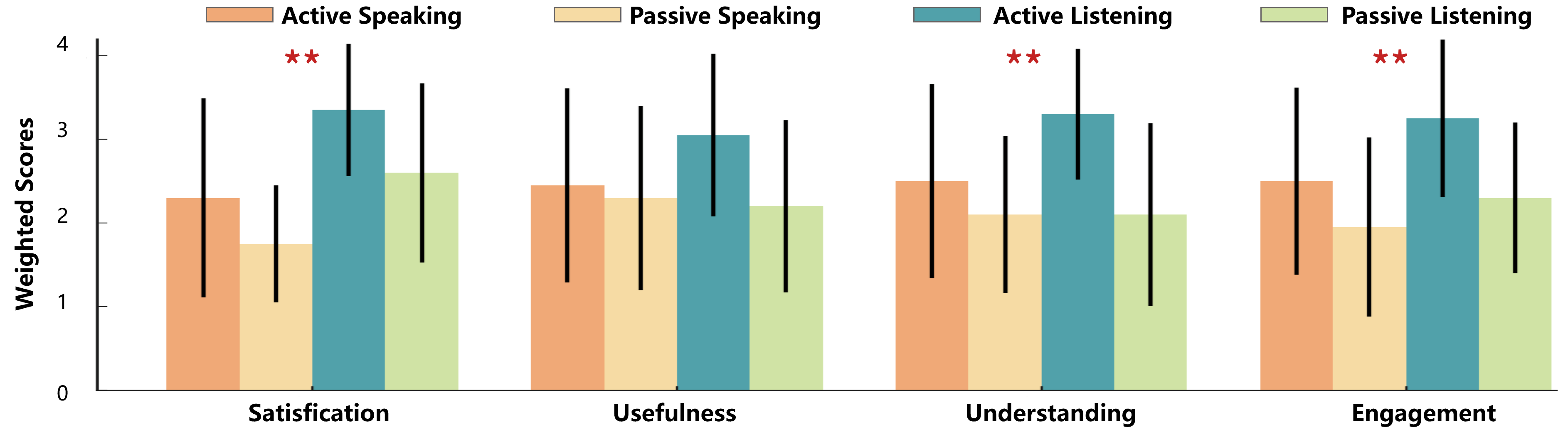}
    \caption{Weighted preference scores for four patterns across four dimensions. Error bars represent weighted standard deviations. Statistical significance is indicated as ** p < 0.01.}
    \label{fig:patterns}
\end{figure}

We first ran Shapiro-Wilk tests and found that the data were not normally distributed, so we used the Mann-Whitney U test for analysis. The results are shown in Fig.~ \ref{fig:res}. To understand participants’ preferences for different conversation patterns, we also asked them to rank four types across four dimensions. Rankings were converted to weighted scores (1st = 4, 2nd = 3, 3rd = 2, 4th = 1), and used the Kruskal-Wallis test. The results are shown in Fig.~\ref{fig:patterns}.

\textbf{Understanding}: Q1 (\emph{I find it helpful for understanding the diverse perspectives on the exhibition more comprehensively.}) showed a significant difference ($U=128.5, p<0.05$), with the \textbf{\textsc{simviews}} ($Md=6.0, IQR=2.0$) was higher than the \textbf{\textsc{base}} ($Md=5.5, IQR=2.0$). Similarly, Q2 (\emph{ I find it helpful for understanding the exhibition more deeply.}) also revealed a significant difference ($U=129.5, p<0.05$), with the \textbf{\textsc{simviews}} ($Md=6.0, IQR=1.0$) outperforming the \textbf{\textsc{base}} ($Md=5.0, IQR=1.75$).

\textbf{Engagement}: Q3 (\emph{I think it improved my engagement with the exhibition}) showed a significant difference ($U=108.5, p<0.05$), the \textbf{\textsc{simviews}} ($Md=6.0, IQR=1.0$) was significantly higher than the \textbf{\textsc{base}}($Md=4.0, IQR=1.75$). However, Q4 (\emph{I am curious about the exhibits and actively explore their details}) did not reach significance ($U=142.0, p=0.099$).

\textbf{Satisfaction}: Q5 (\emph{I feel it met my expectations.}) showed a significant difference ($U=108.0, p<0.01$), with the \textbf{\textsc{simviews}} ($Md=6.0, IQR=1.75$) receiving higher ratings than the \textbf{\textsc{base}} ($Md=5.0, IQR=2.0$). Similarly, Q6 (\emph{I hope to see it in more museums.}) also reached significance ($U=110, p<0.05$), with the \textbf{\textsc{simviews}} ($Md=6.5, IQR=2.0$) again scoring higher than the \textbf{\textsc{base}} ($Md=5.0, IQR=2.0$).

\textbf{Usefulness}: Q7 (\emph{I can easily get diverse perspectives through it.}) showed a significant difference ($U=129, p<0.05$), with participants rating the \textbf{\textsc{simviews}} ($Md=6.0, IQR=1.0$) higher than the \textbf{\textsc{base}} ($Md=5.5, IQR=4.0$). On the other hand, Q8 (\emph{I can get useful information through it}.) showed no significant difference ($U=147, p=0.13$).

As shown in Fig. \ref{fig:patterns}, 
participants consistently preferred \textbf{active listening} over other conversation patterns in \textbf{\textsc{simviews}}. Understanding ($M=3.30$, $SD=0.80$, $p<0.01$), Engagement ($M=3.25$, $SD=0.97$, $p<0.01$), and Satisfaction ($M=3.35$, $SD=0.81$, $p<0.01$) both showed significance. And Usefulness ($M=3.05$, $SD=0.99$, $p=0.076$) showed no significance.

\subsection{Analysis of Conversational Behavior}
Two researchers coded and cross-validated the video data, counting each user-agent exchange as one conversational turn; key metrics are summarized in Table~\ref{table:freq}. Specifically, participants engaged in more conversations in the \textbf{\textsc{simviews}}, with a higher total number of turns ($M=8.1$, $SD=4.15$) compared to the \textbf{\textsc{base}} ($M=3.4$, $SD=2.35$). When focusing on the number of participants actively initiated turns (excluding follow-up questions and passive responses), \textbf{\textsc{simviews}} ($M=5.1, SD=3.13$) also outperformed the \textbf{\textsc{base}} ($M=2.16, SD=1.47$). We also examined whether participants engaged in follow-up turns during the study. The number of follow-up questions was higher in the \textbf{\textsc{simviews}} ($M=2.6,SD=2.3$) compared to the \textbf{\textsc{base}} ($M=0.96, SD=1.47$). Additionally, the maximum number of follow-up turns was also higher in the \textbf{\textsc{simviews}} ($M=1.4, SD=0.94$) than in the \textbf{\textsc{base}} ($M=0.9, SD=1.45$), indicating deeper engagement.

\begin{table}[h]
\caption{Turns of Conversation between \textbf{\textsc{simviews}} and \textbf{\textsc{base}}}
\centering
\begin{tabular}{lcc}
\toprule
\textbf{Measure} & \textbf{\textsc{simviews}} & \textbf{\textsc{base}}  \\
\midrule
Total number of turns & 8.1 (4.15) & 3.4 (2.35) \\
Participants initiating turns & 5.1 (3.13) & 2.16 (1.47) \\
Follow-up turns & 2.6 (2.3) & 0.96 (1.47) \\
Max follow-up turns & 1.4 (0.94) & 0.9 (1.45) \\
\bottomrule
\end{tabular}
\label{table:freq}
\end{table}

As shown in Table~\ref{tab:conv_stats}, we further analyzed the conversation patterns in the  \textbf{\textsc{simviews}}. Participants engaged in an average of 2.4 \textbf{active speaking} ($SD=1.9$), with 12/20 participants initiating at least one follow-up turn ($M=1.4, SD=2.09$), and the maximum number of follow-up turns was 0.9 ($SD=1.02$). In contrast, during \textbf{passive speaking}, 4/20 participants did not respond at all. The overall response rate to agent-initiated prompts was $50\%$, and only 2 participants engaged in any follow-up turns after their initial reply($M=0.25, SD=0.64$). For agent-to-agent conversations,  only 1 participant completely ignores \textbf{passive listening}. 15/20 participants chose to join the discussion after listening, with an average of 2.55 \textbf{active listening} ($SD=1.93$), accounting for 21.25\% of all listening behaviors. Additionally, 10/20 participants engaged in at least one follow-up turns during \textbf{active listening} ($M=0.94, SD=1.15$), and a maximum of 0.65 follow-up turns ($SD=0.75$).

\begin{table}[h]
\centering
\caption{Conversation Pattern Statistics in \textsc{simviews}}
\resizebox{.48\textwidth}{!}{%
\begin{tabular}{lcccc}
\hline
\textbf{Measure} & \makecell{\textbf{Active} \\ \textbf{Speaking}} & \makecell{\textbf{Active} \\ \textbf{Listening}} & \makecell{\textbf{Passive} \\ \textbf{Speaking}} & \makecell{\textbf{Passive} \\ \textbf{Listening}} \\
\hline
\makecell[l]{Total numbers} & 2.4 (1.9) & 2.55 (1.93) & 0.55 (0.69) & -- \\
\makecell[l]{Follow-up turns} & 1.4 (2.09) & 0.94 (1.15) & 0.25 (0.64) & -- \\
\makecell[l]{Max follow-up turns} & 0.9 (1.02) & 0.65 (0.75) & 0.25 (0.64) & -- \\
\hline
\end{tabular}
}
\label{tab:conv_stats}
\end{table}

\subsection{Comprehension Scores}

After participants completed each task, they were asked to take a knowledge test and articulate their understanding of the diverse perspectives presented in the exhibits. The knowledge test included 10 multiple-choice questions per exhibit, derived from literature and refined by a museum expert. Each question was worth 5 points, with a maximum score of 50 per exhibit. As shown in Table \ref{table:understand}. Participants scored 41.5 points ($SD=9.88$) in the \textbf{\textsc{simviews}}, compared to an average score of 43.5 points ($SD=8.44$) in the \textbf{\textsc{base}}.

For the viewpoint statements, researchers first coded three keywords for each perspective, which were then validated and refined with input from museum experts. Then researchers independently coded participants’ statements and scored them as follows: 2 points for a direct keyword match, 1 point for partial relevance, and 0 for no match. Specifically, participants' statements in the \textbf{\textsc{simviews}} received an average score of 3.21 ($SD=1.72$), while those in the \textbf{\textsc{base}} received an average score of 2.43 ($SD=1.40$).

\begin{table}[H]
\caption{Scores of Knowledge Test and Viewpoint Statement}
\centering
\begin{tabular}{lcc}
\hline
\textbf{Measure}        & \textbf{\textsc{simviews}} & \textbf{\textsc{base}} \\ \hline
Knowledge Test & 41.5 (9.88)            & 43.5 (8.44)               \\
Viewpoint Statement   & 3.21 (1.72)            & 2.43 (1.40)             \\ \hline
\end{tabular}
\label{table:understand}
\end{table}

\subsection{Semi-structured Interview}
After transcribing all recordings, two researchers reviewed three randomly selected transcripts and applied an open coding approach, combining deductive and inductive techniques to develop a codebook \cite{braun2012thematic}. They resolved disagreements through discussion, and the research team held regular meetings to refine the codebook until a consensus was reached.

\subsubsection{Tangible and Credible Presentation of Perspectives}

Multiple agents with distinct identities, appearances, and voices made the perspectives more tangible. "\textit{Visually presenting three different viewpoints with three distinct tones of voice}" (P11). This approach clarified and separated perspectives, making them easier to access. \textit{"Having three agents each state their own viewpoint is clearer than having one agent blend all the viewpoints together}" (P2). In contrast, participants found the \textbf{\textsc{base}} more confusing, as all viewpoints came from the same person. "\textit{Sometimes, I couldn’t tell which viewpoint he was referring to}" (P10). Moreover, the single agent reduced the credibility of the perspectives. "\textit{The guide was merely paraphrasing others' views, so I felt their statements might not be professional since they weren’t experts}" (P9). Whereas multi-agents with specialized identities enhanced credibility: "\textit{Since they are professionals in their respective fields, I find their words more trustworthy and am more willing to listen}" (P13).

\subsubsection{Curiosity-driven Proactive Engagement}
Moreover, most participants stated that the agents’ identities attracted them to engage in the \textbf{active speaking} "\textit{Their identities make me curious to ask relevant questions}" (P19). The identity also provided clear guidance to participants: "\textit{If I have any question, I can ask the agent associated with the perspective}" (P3). For example, in the study, P9 actively asked the museum curator: "\textit{I noticed that you are a curator. Performance art is usually seen in art exhibitions rather than in museums. so what do you make of this exhibit?}”. Meanwhile, conflicts between agents also sparked participants' curiosity. "\textit{When I hear a conflict, I want to listen and join in}" (P9). This increased engagement led to more follow-up questions and extended discussions in the \textbf{\textsc{simviews}}. "\textit{Conflicts make me more eager to participate}" (P15). Participants also took clear stances after listening. For instance, P9 tried to persuade the Native American scholar: "\textit{He was trying to convince me, which made me want to keep discussing}". Data analysis showed that \textbf{active listening} was the most favored pattern among participants, with 15/20 choosing to join the conversation while \textbf{passive listening}. 

\subsubsection{Thinking and Memory Enhances Understanding}

Conversations enhanced participants' understanding of diverse perspectives. On the one hand, conversations encouraged participants to think deeply, as P18 said, "\textit{I tried to put myself in their shoes and later understood their viewpoint}". On the other hand, repetition in conversations also reinforced the memory, "\textit{He kept repeating the same thing, which deepened my understanding}" (P11).  In contrast, 6 participants found the \textbf{\textsc{base}} harder to follow, "\textit{He stated viewpoints one after another before I could grasp the previous one}" (P1). Results showed that participants in the \textbf{\textsc{simviews}} ($M=3.21$) scored higher than those in the \textbf{\textsc{base}} ($M=2.43$), suggesting they remembered and understood more diverse perspectives in \textbf{\textsc{simviews}}. 

However, if participants lacked patience and engagement, they might miss key information, which could explain why the \textbf{\textsc{simviews}} scored slightly lower in some knowledge tests. As P8 noted, "\textit{I should have followed up, then I would have known the answer}".

\section{DISCUSSION}
\subsection{Presenting Perspectives via Multiple Agents}
\subsubsection{Clarifying Diverse Perspectives through Multiple Agent and Layered Presentation}
Diverse museums present extensive information and varied viewpoints, which can easily overwhelm visitors, especially when multiple perspectives are from a single source. \textbf{\textsc{simviews}} addresses this challenge by distributing viewpoints across multiple agents, each with a unique professional identity and multimodal features. It allows users to associate each perspective with a specific agent, reducing confusion and enhancing clarity, as P3 noted: “\textit{If I have any question, I can ask the agent associated with the perspective}”. Prior work also supports the effectiveness of multi-agent systems in conveying contrasting views and aiding comprehension \cite{Kantharaju2018Two}. Visual and behavioral differences among agents further reinforce the clarity and memorability of the content \cite{yasrebi2020visualizing}.

However, some users felt that multi-agent conversations lacked essential background information, creating a barrier to understanding. This may explain why knowledge test scores in \textsc{\textbf{simviews}} were slightly lower than in \textsc{\textbf{base}}. They recommended introducing a guide agent to offer basic exhibit information before the perspective-based conversations begin. This layered structure supports users’ understanding of perspectives while reducing cognitive load \cite{rzayev2019fostering}, and aligns with the principles of Learning Hierarchies \cite{white1973research}. Future virtual museums should continue to explore such hierarchical strategies for presenting perspectives clearly and gradually.

\subsubsection{Facilitating Proactive Questioning through Characterized Agents}
It remains a challenge to present diverse perspectives appropriately while attracting and sustaining a visitor's attention. In \textsc{\textbf{simviews}}, we assign professional identities to agents to make conversations feel more purposeful and credible, encouraging engagement \cite{Sharp2020Impact, Kantharaju2018Two}. Our results also showed that participants in \textsc{\textbf{simviews}} engaged more actively ($M=8.10$) than \textsc{\textbf{base}} ($M=3.40$). Because participants expressed a greater willingness to interact with characterized visitor agents who appeared more professional than with guide agents who seemed to provide only surface-level information. Participants valued the agents’ professional fields  (e.g., whether a biologist specialized in tigers or lions)  for targeted questioning, rather than detailed identity attributes (e.g., age or name). Research also confirms that expertise boosts agent persuasiveness \cite{Mahmood2022Effects}. 

Meanwhile, participants expressed a desire for more identity diversity among agents to access richer perspectives \cite{sawyer2013studying}. However, some noted that increasing agent variety still couldn't fully capture all perspectives. They suggested tailoring agent recommendations based on user preferences. Prior work has explored using LLMs to personalize museum content or visitor paths \cite{trichopoulos2023large2, trichopoulos2023large, liu2024toward}, pointing to future opportunities for LLM-driven systems to recommend professional perspectives aligned with individual interests.

\subsection{Conversations for Understanding  and Engagement in Diverse Museums}

\subsubsection{Deepening Understanding through Discussion and Conversational Feedback}
Diverse perspectives increase the complexity of museum content, making it harder to understand. To support comprehension, we designed visitor-to-visitor conversations where users can gain diverse perspectives. Participants reported a deeper understanding through these conversations, with results indicating that \textbf{active listening} ($M=3.30$)  was more helpful than \textbf{active speaking} ($M = 2.50$) for understanding perspectives. This was because the discussion prompted memory and thinking more effectively than a single-agent interpretation. As P11 noted, “\textit{He repeatedly stated the same point, and this repetition actually deepened my memory.}” Others described how the discussion helped them consider alternative viewpoints, as P18 shared, “\textit{I try to think from his perspective, so later I was able to better understand his viewpoint.}” The viewpoints presented through discussion are expressed and reinforced, thus enhancing the memorability and acceptance of the information. 

However, participants noted that the conversational multimodal feedback was limited. They suggested that agents should better simulate physical museum behavior by referencing exhibits through gestures and movement to support understanding \cite{ho2025enhancing, hornecker2022human}. For example, when the ethicist discussed human remains in Lion Attacking a Dromedary, participants expected the agent to walk toward the exhibit and visually direct their attention. Prior work supports this need, showing that synchronized multimodal cues (e.g., text highlights or pointing gestures) enhance comprehension in complex environments \cite{wang2024virtuwander}. Improving such feedback could further support users’ understanding of diverse perspectives.

\subsubsection{Enhancing Engagement through Agent Debate and User-Controlled Conversations}

Agent debate serves as an effective mechanism for sustaining visitor engagement in virtual museums. Disagreements between visitor agents raised participants' curiosity, fulfilling their desire for "\textit{watching a bustling scene}" and attracting them to actively participate. Some participants noted, "\textit{When I hear a conflict in their discussion, I want to listen, and then I want to join in}". Results also showed that 15/20 participants transitioned from \textbf{passive listening} to \textbf{active listening} during listening discussions. Such conflicts not only sparked participants' interest but also prompted them to continue questioning and engaging in deeper discussions. We observed that the number of follow-up turns in \textbf{\textsc{simviews}} ($M=2.60$) was higher than in \textbf{\textsc{base}} ($M=0.96$), and \textbf{active listening} had a relatively high number of follow-up turns($M=0.94$), suggesting that conflict discussions effectively enhance engagement \cite{mivsurac2012advantages}.

In addition, participants suggested leveraging the advantages of virtual museums over physical museums to enhance the user's dominant position \cite{li2019appropriate}.  They favored and engaged in \textbf{active speaking}, which had the highest number of follow-up turns ($M=1.4$), while \textbf{passive speaking} ($M=0.25$) had the lowest. Although some expressed concerns about social pressure while speaking, most felt the virtual setting reduced anxiety \cite{kampmann2016exposure, zamanifard2023social}. Some also hoped to assign a visitor agent to follow them and participate in other conversations together, rather than only joining existing ones. Prior work has explored using natural language in Unity for such control \cite{ito2024demo}, and future research can further expand the conversation pattern and enhance the user's dominant position in virtual museums.

\subsection{Limitations and Future Work}
Our study has some limitations. First, the exhibits may not have interested all participants, potentially affecting their engagement and overall evaluation. Future research could include a broader range of exhibit themes to attract diverse participants and improve generalizability. Additionally, this study was conducted in a simplified virtual museum without VR devices, which could affect user interaction and engagement \cite{hurst2016complementing}. Future studies could explore VR environments to see if multi-agent systems offer a more immersive experience. Lastly, interruptions can effectively enhance the fluency and engagement of conversations. However, our design currently lacks support for interrupting conversations, which could be considered for future implementation. These approaches would lead to a richer, more personalized user experience in virtual museums.
\section{CONCLUSION}
Diverse perspectives enrich museum experiences, yet most virtual museums still rely on singular, authoritative narratives. Using recent advances in LLMs, we developed SimViews, a multi-agent system that presents diverse perspectives through simulated visitor-to-visitor conversation. Evaluation results show that SimViews effectively supports understanding of diverse perspectives and increases user engagement. This highlights the promise of multi-agent conversation for building more inclusive and diverse virtual museum experiences. We hope our insights will inspire future efforts toward more diverse, equitable, and inclusive virtual museum design.

\begin{acks}
This work is partially supported by Guangzhou Higher Education Teaching Quality and Teaching Reform Project (No. 2024YBJG070),  HKUST Practice Research (No. HKUST(GZ)-ROP2025002), Guangzhou-HKUST(GZ) Joint Funding Project (No. 2024A03J0617), Education Bureau of Guangzhou Municipality Funding Project  (No. 2024312152), Guangdong Provincial Key Lab of Integrated Communication, Sensing and Computation for Ubiquitous Internet of Things (No. 2023B1212 010007), the Project of DEGP (No.2023KCXTD042), and Artificial Intelligence Research and Learning Base of Urban Culture (No. 2023WZJD008). 
\end{acks}

\bibliographystyle{ACM-Reference-Format}
\bibliography{sample-base}
\end{document}